\documentclass[twocolumn,showpacs,preprintnumbers,amsmath,amssymb]{revtex4-1}
\usepackage{epsfig}
\usepackage{graphicx}
\usepackage{epsfig}
\usepackage{rotating}
\usepackage{amssymb}
\usepackage{amsmath}
\usepackage{graphicx}
\usepackage{midfloat}

\usepackage{xcolor}
\usepackage{pseudocode}
 \expandafter\let\csname equation*\endcsname\relax
  \expandafter\let\csname endequation*\endcsname\relax
   
\newcommand{\vm}[1]{\boldsymbol{#1}}

\begin{document}

\title{Statistical properties of interaction parameter estimates in direct coupling analysis}

\author{Yingying Xu}
\affiliation{Department of Computer Science, School of Science,
Aalto University, P.O.Box 15400, FI-00076 Aalto, FINLAND}
\altaffiliation{Email address: yingying.xu@aalto.fi}

\author{Erik Aurell}
\affiliation{Department of Computational Science and Technology, KTH-Royal
Institute of Technology, SE-100 44 Stockholm, Sweden }
\affiliation{Depts of Applied Physics and Computer Science, Aalto University, FIN-00076 Aalto,
Finland}
\affiliation{Institute of Theoretical Physics, Chinese Academy of Sciences, Beijing, China}

\author{Jukka Corander}
\affiliation{Department of Mathematics and Statistics, University of Helsinki, 00014 Helsinki, Finland;
Department of Biostatistics, University of Oslo, 0317 Oslo, Norway
}

\author{Yoshiyuki Kabashima}
\affiliation{Department of Mathematical and Computing Science, School of Computing, Tokyo Institute of Technology, Tokyo 152-8552, Japan}

\begin{abstract}
We consider the statistical properties of interaction parameter estimates obtained by the direct coupling analysis (DCA) approach to learning interactions from large data sets. Assuming that the data are generated from a random background distribution, we determine the distribution of inferred interactions. Two inference methods are considered: the $L_{2}$ regularized naive mean-field inference procedure (regularized least squares, RLS), and the pseudo-likelihood maximization (plmDCA). 
For RLS we also study a model where the data matrix elements are real numbers, identically and independently generated from a Gaussian distribution; in this setting we analytically find that the distribution of the inferred interactions is Gaussian.
For data of Boolean type, more realistic in practice, the inferred interactions do not generally follow a Gaussian. 
However, extensive numerical simulations indicate that their distribution can be characterized by a single function determined 
by a few system parameters after normalization by the standard deviation. This
property holds for both RLS and plmDCA and may be exploitable for inferring the distribution of extremely large interactions from 
simulations for smaller system sizes.  
\end{abstract}

\pacs{}
\maketitle

{\bf Introduction}---Identifying meaningful pairwise relationships between entities from very high-dimensional data is
a central task in data science~\cite{WainwrightJordan2008}. 
Of particular interest is the family of methods now generally known as direct coupling analysis (DCA), which aims
to characterize such relationship by the parameters of a Potts model or an Ising model inferred from the data.
Important progress has been made on inferring residue-residue contacts in proteins from
multiple sequence alignments (MSA) of many homologous proteins~\cite{WeigtWhite2009,Burger2010,morcos2011},
and has led to a breakthrough in \textit{in silico} protein structure prediction~\cite{Marks2012-NatureBio,Baker2017}.
Among other applications we point to a method recently developed by us to detect of epistatic interactions in bacteria
from population-wide dense sequencing data~\cite{Meala_plmDCA}. 
The area was reviewed from the methodological point of view at an early stage in~\cite{Roudi2009} and more recently in~\cite{SteinMarksSander2015} and~\cite{NguyenBergZecchina}.

The above mentioned methods and applications share several challenging aspects. The first,
discussed in detail in \cite{NguyenBergZecchina} 
and~\cite{montanari2015} (and elsewhere),
is that learning non-trivial probabilistic
models from large data sets is computationally hard; in practice it can only be done approximately or relying on learning
criteria weaker than maximum likelihood. The second is that the relevant data typically have fewer samples ($n$) than 
the number of parameters characterizing the model ($p$). For the protein contact prediction problems $n$ is
often in the range $10^3-10^5$ while the number of parameters in the corresponding Potts models
grows quadratically with the length of the protein, and can hence realistically be $10^6-10^7$. Inference schemes therefore need to
be regularized, as will be exemplified below by the RLS and plmDCA algorithms. 
The third aspect is that while at an intermediate stage one infers many more parameters than there is data,
the final goal is to provide only (relatively) a small number $k$ of predictions: in all successful applications 
to date $k\ll p$, and usually $k$ has also been smaller (or much smaller) than $n$.

From all the above follows that the important task of assessing the statistical properties and significance thresholds of DCA 
is highly non-trivial.
Indeed, in applications one would like to assign a p-value of a list of, say, $k$ largest predictions.
Alternatively, given an MSA representing some data one would like to be able to say that 
$k$ pairwise relationships can be predicted from this data with some pre-determined level of confidence.
Usually neither of the above has been done, but instead
inference algorithms have been
evaluated by comparing to a ``ground truth'' deduced from separate experimental data, in particular, for the contact
prediction problem, by comparing to
protein crystal structures \cite{WeigtWhite2009,morcos2011,Magnus2013,Magnus2014,RLS2014,SteinMarksSander2015}. 
For a majority of potential future applications such a ground truth will not be available and there is therefore a need to 
develop statistical tests of the results of the inference. 
In fact the only previous step in this direction that we are aware of is our recent use of extreme value distribution theory (Gumbel distribution) to describe the background distribution for a variant of DCA where only 
the exceptionally large sampled predictions  
are retained in an intermediate step \cite{Meala_plmDCA}, and which is therefore a kind of special case.
 
In this paper we introduce the systematic study of 
the background and statistical significance of DCA predictions. We point out that the question of statistical significance of DCA is a large deviation problem for a nonlinear transformation of a data matrix, and discuss several examples which can be understood numerically and/or analytically. 

{\bf Naive mean-field inference of Ising models and Regularized Least Squares}---
Consider a data matrix consisting of Boolean variables $\sigma_i^{(r)}$ taking values $\pm 1$ and let the inference task 
be to estimate the interaction parameters (``$J$") in an Ising model
\begin{equation}
P(\underline{\vm{\sigma}})=\frac{1}{Z}\exp\left(\sum_i h_i \sigma_i +\sum_{i\neq j}J_{ij}\sigma_i\sigma_j\right)
\label{Ising model}
\end{equation}
from this data.
The simplest version of variational inference is ``naive mean-field"~\cite{WainwrightJordan2008} where 
$\vm{J}=-\vm{C}^{-1}$ and $C_{ij}=\left \langle \sigma_i\sigma_j\right \rangle- \left \langle \sigma_i \right \rangle 
\left \langle \sigma_j \right \rangle$ is the covariance matrix. Note that
the indices of $J_{ij}$ are defined to be different and that therefore only the off-diagonal elements of the inverse
covariance matrix $\vm{C}^{-1}$ are used.
When $n<L$ naive mean-field is not a well-defined procedure since the covariance matrix then does not have full rank,
and hence has to be regularized by the use of pseudo-counts, as in~\cite{morcos2011}, or a sparsity-promoting $L_1$ regularizer, as in~\cite{jones2012}. 

In the same family Regularized Least Squares (RLS) is an $L_2$-regularized inference method and given by the simple matrix equation
 \begin{equation}
 \label{eq:RLS}
\vm{J}^{\mathrm{RLS}}= - \vm{C} \left(\eta\vm{1} + \vm{C}^2\right)^{-1}
\end{equation}
where $\vm{1}$ is the identity matrix and $\eta>0$ is a positive regularization parameter. Although arguably one of the simplest non-trivial inference procedures one can imagine, to the best of our knowledge RLS was only introduced in the context of DCA in \cite{RLS2014}, and has not been investigated since in the subsequent DCA literature. We will use it here since the matrix form  Eq.~(\ref{eq:RLS}) renders it quite convenient for our purposes.

\begin{figure}[h]
\centering
\includegraphics[width=8cm, height=6cm]{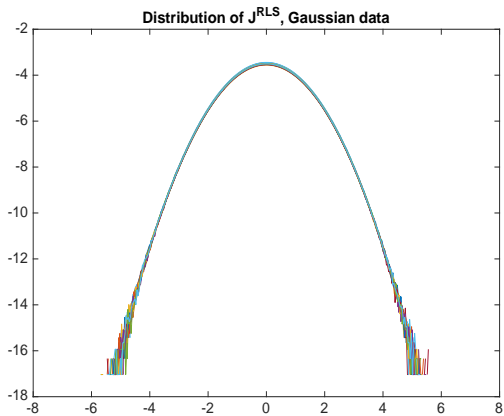}
\caption{Log-histogram of inferred off-diagonal interactions $\vm{J}^{\mathrm{RLS}}$ using RLS algorithm (Eq.~(\protect\ref{eq:RLS})
starting from ${\mathcal N}(0,1)$ i.i.d. Gaussian data, $\eta=0.1$.
Number of loci in this series was $L=10000$ and number of
samples $n=1000, 2000,\ldots, 20000$. 
All distributions scaled by the standard deviation of each instance.
}
\label{GaussianCase2Log}
\end{figure}

{\bf A solvable example}---
A model problem of RLS inference can be 
completely understood by random matrix theory. 
Assume that the elements in the data matrix $X\in\mathbb{R}^{n\times L}$ are real and Gaussian distributed ${\cal N}(0,\sigma^2)$. 
The covariance matrix $\vm{C}=\frac{1}{n}X^\top X$ is then a Wishart matrix. 
The spectrum of $\vm{C}$ converges by the Marcenko-Pastur law almost surely to a limit when $n$ and $L$ tend simultaneously to infinity, and since $\vm{J}^{\mathrm{RLS}}$ is related to $\vm{C}$ by (\ref{eq:RLS}) the spectrum of $\vm{J}^{\mathrm{RLS}}$ is almost surely a non-linear transformation of the Marcenko-Pastur distribution.
Furthermore, the distribution of the individual elements of the distribution of the individual elements of $\vm{J}^{\mathrm{RLS}}$
can be obtained from a singular value decomposition of matrix $X$,
\begin{equation}
X=USV^\top,
\end{equation}
and regarding the left and right eigen-bases as samples from the uniform distributions of orthogonal matrices. 
The covariance matrix can then be expressed as 
\begin{eqnarray}
\vm{C}&=&\frac{1}{n}VS^{2}V^\top\\\nonumber
&=&V\Lambda V^\top\\\nonumber
&=&\left(\sum_{k=1}^{L}u_{ik}\lambda_{k}u_{jk}\right),
\end{eqnarray}
where $\lambda_{k}$ is the $k$th eigenvalue of covariance matrix $\vm{C}$ and $u_{ik}$ is the $k$th elements of $i$th eigen-base of matrix $\vm{C}$.
From $(\ref{eq:RLS})$ the inferred interaction matrix obtained by RLS is
\begin{eqnarray}
\left(J^{\mathrm{RLS}}_{ij}\right)=\left(\sum_{k=1}^{L}\frac{\lambda_{k}}{\eta+\lambda_{k}^2}u_{ik}u_{jk}\right).
\end{eqnarray}
In this situation $u_{ik}$ are samples from the uniform distributions of orthogonal matrices, and when the dimension of the matrix $\vm{C}$ goes to infinity 
$u_{ik}$ can be handled as random numbers \cite{KabaCDMA} that satisfy
\begin{eqnarray}
\overline{u_{ik}}=0, \overline{u_{ik}u_{jl}}=\frac{1}{L}\delta_{ij}\delta_{kl}.
\label{eq:u_mean}
\end{eqnarray}
Condition $(\ref{eq:u_mean})$ means that each diagonal component converges to an $\mathcal{O}(1)$ constant as 
\begin{eqnarray}
J^{\mathrm{RLS}}_{ii}
&=&\langle \frac{\lambda}{\eta+\lambda^2} \rangle,
\end{eqnarray}
where the brackets $\langle\cdot\rangle$ denote the expectation with respect to eigenvalue distribution $\rho(\lambda)$ of the covariance matrix $\vm{C}$.
The off-diagonal elements similarly follow a zero mean Gaussian distribution
\begin{equation}
J^{\mathrm{RLS}}_{ij} \sim \mathcal{N}(0,v_{J}^2)
\end{equation}
with variance
\begin{equation}
v_{J}^2\simeq\frac{1}{L-1}\left(\langle\left(\frac{\lambda}{\eta+\lambda^2}\right)^{2}\rangle-\langle\frac{\lambda}{\eta+\lambda^2}\rangle^2\right).
\label{estimated_offdiagonal_variance}
\end{equation}
To test the above theory 
we generated standard i.i.d Gaussian random variables as data matrix elements,
representative results summarized in Fig.~\ref{GaussianCase2Log}.
This indicates that after standardization $J_{ij}\leftarrow (J_{ij}-\overline{J})/v_{J}$, 
where the mean $\overline{J}$ vanishes in the current case, the
coupling distribution collapses to a {\em single function} of $\mathcal{N}(0,1)$ 
irrespectively of any system parameters.

\begin{figure}[h]
\centering
\includegraphics[width=8cm, height=6cm]{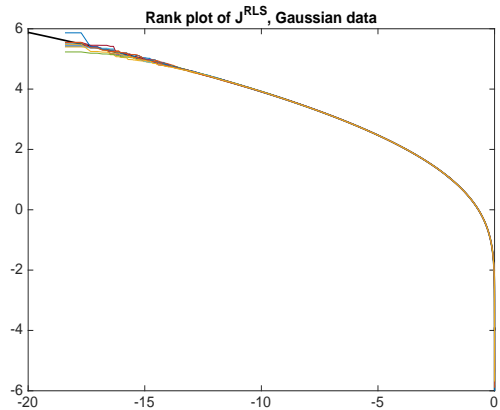}
\caption{Rank plot of inferred off-diagonal couplings.
The $J$'s are centered and scaled by the standard deviation 
of each instance, and then
re-ordered as $J^{1}\geq J^{2}\geq\cdots J^{r}\geq\cdots$ where $r$ indicates the rank.
Data lines (colour online): $J^{r}$ vs. $\log(r/N_L)$ where $N_L=L(L-1)$ is the total number of inferred off-diagonal couplings.
Black curve: $x=\log \frac{1}{\sqrt{2\pi}}\int_y^{\infty}e^{-\frac{1}{2}t^2}\,dt$.
}
\label{RankPlotGaussian}
\end{figure}

Agreement between numerics and the theoretical prediction is good.
However, as discussed above 
DCA works by selecting only the exceptionally large predictions and what would matter for an evaluation of the
statistical background is the distribution over such rare events.
The deviation in the tail (or any finite range) between an empirical and a predicted probability distribution
can be quantified by \textit{e.g.} an Anderson-Darling test~\cite{Anderson-Darling},
but to connect to current practice in the DCA field we instead display in Fig.~\ref{RankPlotGaussian} the
largest predicted couplings in a rank plot.
Clearly the empirical distribution is quite close to Gaussian also according to this more challenging test.     

{\bf Boolean data, plmDCA: finite-size scaling property}---
We turn to RLS-inferred couplings from random Boolean data. 
As in realistic examples the bias of each variable in each position $i$ varies, we examined the case where the bias 
$f_i={\rm Pr}(\sigma_i=+1)=1-{\rm Pr}(\sigma_i=-1)$ is uniformly distributed on region $(0,1)$. 
The obtained coupling distribution is neither a 
Gaussian nor another analytically expressible distribution. 
Nevertheless, the experimental results indicate that as $L$ grows keeping aspect ratio $\alpha=n/L$ and 
$\eta$ fixed, the coupling distribution after standardization 
collapses to a single function characterized by $\alpha$ and $\eta$. 
In analogy with well-established use in percolation theory and
constraint satisfaction we term this a {\em finite-size scaling property}~\cite{Kirkpatrick1297}.

\begin{figure}[h]
\centering
\includegraphics[width=8cm, height=6cm]{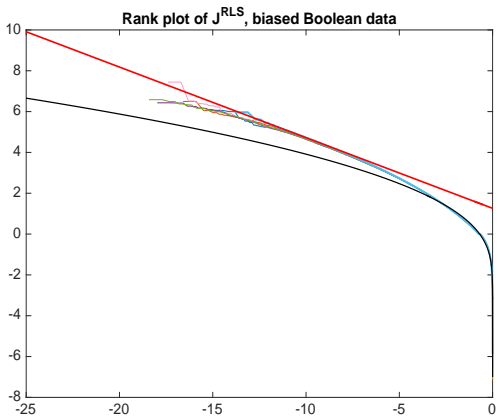}
\caption{Rank plot of RLS algorithm inferred off-diagonal couplings from random Boolean data generated as the bias of each variable in each position $f_i$ were uniformly distributed on region $(0,1)$.
In this series, the aspect ratio $n/L=0.1$ and the regularization parameter $\eta=0.5$. The thick light blue curve was the average of  $100$ experiments for random data of loci size $L=1000$, which is a smooth curve. The straight line
was extrapolated by the very tail $100$ points (about top $0.01\%$ rank data) of this light blue curve.
Number of loci in the colour curves which has a vibrating tail were $L=2000, 4000, \ldots, 10000$. The black curve is drawn by standard Gaussian as a refernce line. All curves scaled by the standard deviation of each instance. }
\label{RankPlot_unibias_RLS}
\end{figure}

This may be exploitable for practical purposes. 
As RLS needs the matrix inversion operation, experimental evaluation of the background distribution becomes
computationally infeasible as $L$ very large. 
However, the scaling property allows us to infer the distribution of large $L$ via Monte Carlo assessments for smaller $L$. 
A crucial drawback of this approach is that the information for 
the extremely large predictions that occur with a probability smaller than 
$2/(L(L-1))$ cannot be evaluated accurately. However, the empirical observation 
that the tails of the normalized coupling distribution typically go down 
like a Gaussian (Fig.\ref{RankPlot_unibias_RLS}) indicates that the straight line extrapolated from the very tail
in the rank plot acts as an upper bound for the extremely rare predictions, 
which can be useful for screening relevant couplings from the background.

We have thus far analyzed the RLS algorithm, which is based on mean field assumption, because of the simplicity of the resulting mathematical and numerical analysis. Another approximation algorithm for estimating the couplings in (\ref{Ising model}) is pseudo-likelihood maximization (plmDCA)~\cite{Magnus2014}, which has been shown to have very accurate performance in predicting the residue contacts in protein structures. 
plmDCA maximizes probability of each element $\sigma_i$ conditioned by all other elements $\vm{\sigma}_{\backslash i}$, $P(\sigma_{i}|\vm{\sigma}_{\backslash i})$, and regularizes the couplings $\vm{J}$ and external fields $\vm{h}$ by the $L_2$ norm. 
For details, see Ref. \cite{Magnus2014}. 
\begin{figure}[h]
\centering
\includegraphics[width=8cm, height=6cm]{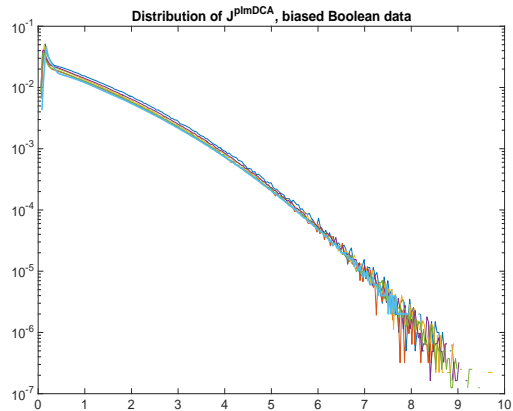}
\caption{Log-histogram of plmDCA algorithm inferred off-diagonal interactions from random Boolean data generated as $f_i$ were uniformly distributed on region $(0,1)$. In the experiments, $n/L=0.1$ and $\lambda_{J}=\lambda_{h}=0.5$. The thick light blue curve was the average of  $100$ experiments for random data of loci size $L=1000$, which is a smooth curve in low ranking region. Number of loci in the colour curves which has a vibrating tail were $L=2000, 2500, \ldots, 4000$. 
All distributions scaled by the standard deviation of each instance.}
\label{Log_ProbDens_plmDCA_unibias}
\end{figure}
\begin{figure}[h]
\centering
\includegraphics[width=8cm, height=6cm]{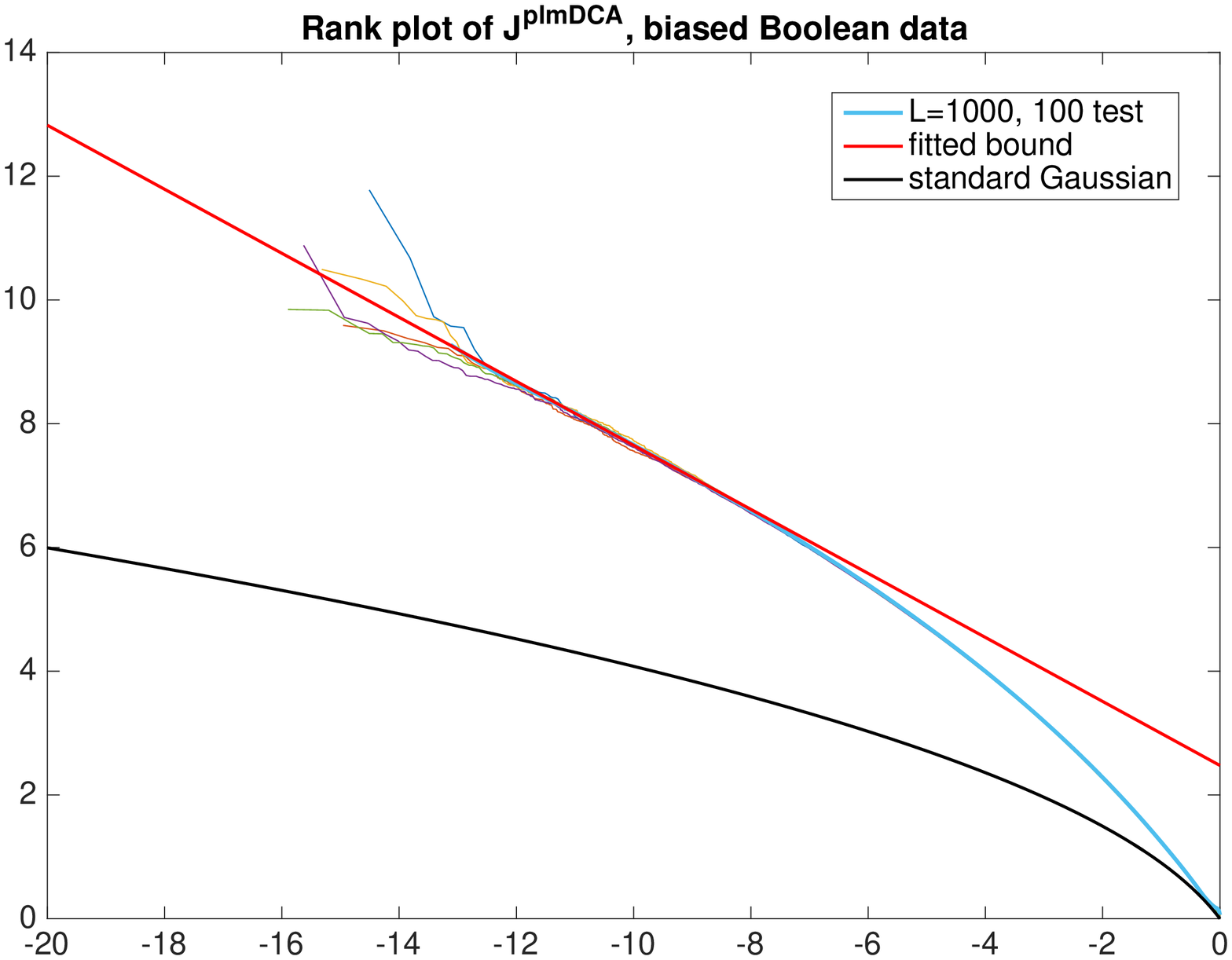}
\caption{Rank plot of plmDCA algorithm inferred off-diagonal couplings from random Boolean data same as in Fig.\ref{Log_ProbDens_plmDCA_unibias} with the same parameter settings. The thick light blue curve was the average of  $100$ experiments for random data of loci size $L=1000$, which is a smooth curve. The straight line
was extrapolated by the very tail $100$ points of this light blue curve. Other single test with different size $L$ are fluctuated around the averaged curve in the tail. Black curve: standard Gaussian. All curves scaled by the standard deviation of each instance.}
\label{rank_plot_plmDCA_unibias_randomdata}
\end{figure}
Here we examine the coupling distribution for plmDCA when the input  Boolean data follows the uniform bias distribution. The result is shown in Fig.\ref{Log_ProbDens_plmDCA_unibias}. Since Frobenius norm was taken on $||\vm{J}_{ij}(a,b)||_{F}$ for each $i,j$ pairs, the output coupling value only has positive values.

Unlike RLS, the relation between the coupling obtained by plmDCA, $\vm{J}^{\rm plmDCA}$, 
and $\vm{C}$ is highly non-trivial, and cannot be analytically expressed. 
However, the experimental results (Fig.\ref{rank_plot_plmDCA_unibias_randomdata}) show that the scaling property holds for plmDCA as well, 
and can be used for inferring the distribution of extremely large couplings 
from extensive simulations for smaller system sizes. 
The value of this property may be more significant for plmDCA than for RLS since plmDCA is computationally demanding and practically difficult to repeat random simulations of large systems many times 
for accurately assessing the background effect.

{\bf Discussion}---The study of the backgound distribution of inferred DCA couplings is of importance for supporting the selection of significant predictions. Our work in this paper proves that 
when the data matrix elements are i.i.d Gaussian variables the distribution of the inferred interactions is also Gaussian.
For data of Boolean type, extensive numerical simulations indicate that their distribution can be characterized by a single function determined 
by a few system parameters (aspect ratio $\alpha$, regularization parameter and column bias $\{f_i\}$) after normalization by the standard deviation; this holds for both RLS and plmDCA. 
This property may be exploitable for inferring the distribution of extremely large interactions from 
simulations for smaller system sizes.

Also, investigations on other background models, for example, a neutral model, could be helpful to separate clonal inheritance\cite{clonal_inheritance} from epistasis\cite{epistasis} as a mechanism for linkage disequilibrium\cite{LD}. In our recent application of DCA to detect epistasis from genome sequence data\cite{Meala_plmDCA}, taking such biological effects into account in the background model could lead to a more accurate method to predict significant couplings. 

\section{Acknowledgements}
This research is supported by the Swedish Science Council through grant 621-2012-2982(EA) and by the Academy of Finland through its Center of Excellence COIN (YX, EA and JC), and JSPS KAKENHI Nos. 25120013 and 17H00764 (YK). 
EA acknowledges support from Chinese Academy of Sciences CAS President's International Fellowship Initiative (PIFI) GRANT No. 2016VMA002.

\bibliography{DCA_background_arXiv}

\end{document}